\documentclass{article}
\usepackage{emulateapj}

\slugcomment{To appear in ApJL}

\begin{document}
\title{Convective Wavelength Shifts in the Spectra of Late-Type Stars}

\author{Carlos Allende Prieto, David L. Lambert, Robert G. Tull, 
and Phillip J. MacQueen}
\affil{McDonald Observatory and Department of Astronomy \\ University of Texas
  \\ RLM 15.308, Austin, TX 78712-1083, \\ USA}
  

\begin{abstract}

We present  ultra-high resolution  spectra for a set of
nearby F-G-K stars on, or close to, the main sequence. The wavelength shifts
 of stellar lines relative to their laboratory wavelengths 
 are measured for more than a thousand Fe {\sc i} lines per star, 
 finding a clear correlation
 with line depth. The observed patterns are  interpreted as
 convective blue-shifts that become more prominent for weaker lines,
 which are formed in deeper atmospheric layers. A morphological 
 sequence with spectral type
 or effective temperature is apparent.
 Two K giant stars have also been studied. The velocity span between 
weak and strong lines  for these stars is larger  than for the 
 dwarfs and subgiants of similar spectral types.
 Our results show that  convective 
  wavelength shifts may seriously compromise the accuracy of 
  absolute spectroscopic radial velocities, but  that an empirical 
  correction may be applied to measured velocities. 
  

\end{abstract}

\section{Introduction}

Convection in late-type stars penetrates into the photosphere, 
producing inhomogeneities. Temperature variations 
of up to a few hundred Kelvin 
are apparent in optical images of the Sun, 
and a velocity field of  several kilometers per second 
is observed in spatially resolved 
 solar spectra. The 
warmer upflows appear as bright granules surrounded by narrower, cooler
downflows. While present technology cannot  resolve stellar granulation,
there is evidence  of its presence in
all late-type stars. The  
temperature and velocity inhomogeneities produce absorption
line profiles which are shifted and asymmetric, effects 
that are readily noticed in  ultra-high dispersion stellar spectra.

An extensive literature exists on solar line asymmetries and shifts
(see, e.g., Neckel \& Labs 1990; Asplund et al. 2000; Pierce \& Lopresto 2000).
Modern stellar studies were pioneered by Dravins (1974, 1985) 
and Gray (1980, 1981). Dravins (1999) provides a short review of recent work.
Line asymmetries and line shifts are consequences of the same
phenomenon and, therefore, give largely equivalent information. When the
number of lines measured is limited, but the available features are mostly
unblended, line asymmetries carry  more information. On the contrary, if
the spectrum is heavily crowded, line shifts, which can be measured
for many lines, are likely to provide 
 more information.
A limited wavelength coverage, largely dictated by the detector size and
the scarcity of high-accuracy laboratory wavelengths, steered stellar 
studies toward the analysis of line asymmetries.

Line bisectors -- a measure of the line asymmetry -- vary smoothly
 across the HR 
diagram (Gray \& Toner 1986; Dravins 1987; Gray \& Nagel 1989; 
Dravins \& Nordlund 1990). Other parameters, such as
rotation, chemical composition, magnetic fields, or binarity, are likely to
play a role, but the limited data available have prevented an in-depth study.
A recent comprehensive work on the Fe {\sc i}
 spectrum  (Nave et al.
1994) has provided accurate laboratory wavelengths for 9501 lines. 
These data and 
 improvements in astronomical instrumentation make it feasible and practical 
 to turn to line shifts
as a complement to line bisectors. 

Gravitational and convective wavelength shifts systematically affect absolute 
determinations of radial velocities.
 Other effects can also introduce systematic radial velocity 
errors, but they are expected to be  less important for most stars 
(Lindegren, Dravins, \&
Madsen 1999). Gravitational redshifts are proportional
to the mass-to-radius ratio. For spectral lines formed in the photospheres of
dwarf stars, the gravitational shift is the same for all lines, 
in the range between 0.4 and 2 km s$^{-1}$. 
When accurate parallaxes and photometry
are available, comparison with evolutionary models allows us to determine dwarf
masses and radii within 8 \% and 6 \% , respectively 
(Allende Prieto \& Lambert 1999).
It is therefore possible to estimate the gravitational shift
within 10 \%, yielding worst-case uncertainties of $\sim 0.2$ km s$^{-1}$.
Convective line shifts may vary with spectral type. 
The velocity difference between weak and strong lines is about
 $\sim 0.6$ km s$^{-1}$ for the Sun (Allende Prieto \& Garc\'{\i}a L\'opez
1998) and probably more than 1 km s$^{-1}$ for the F5 subgiant  $\alpha$ CMi
(Procyon; Allende Prieto et al. 2002), but  no systematic study across
the HR diagram has been published. Obviously, accurate spectroscopic studies of
absolute radial velocities 
cannot afford to  neglect these shifts. 

In this paper, we analyze optical 
spectra of several late-type stars  
obtained with the High Resolution Spectrograph (HRS) coupled to the Hobby-Eberly
Telescope (HET). The high quality and large spectral coverage of the spectra
allow us to measure line shifts for a large number of 
Fe {\sc i} lines. Section \ref{obs} describes the observations and the data. In 
Section \ref{doit} we  describe the analysis,
 and in \S \ref{end} we discuss the results.

\section{Observations}
\label{obs}

Our observations were obtained with the 9.2m HET at 
McDonald Observatory and the HRS (Tull 1998).
The HRS saw first light on  March 17, 2001, starting science observations
shortly afterwards. This fiber-coupled echelle spectrograph uses an R-4
echelle, cross-dispersing gratings, and a mosaic of two 2k$\times$4k CCDs.
 In the selected mode, HRS provides a 
FWHM resolving power 
$R \simeq 120,000$ and complete spectral coverage between 
4094 and 7890 \AA, with the exception
of the  gap (5977--5999 \AA)  between the CCDs.
The observations were performed in {\it queue} mode in May and 
June 2001 (see Table 1). An  observation consisted of two 
exposures that were extracted and combined. 
The final signal-to-noise (S/N) per pixel was $\ge 400$ at most wavelengths.

The spectra were processed with the tasks in the IRAF {\it echelle} package. 
We removed the bias signal, flatfielded and extracted 
the spectra,  subtracting the low background  in the wide 
interorder regions. 
In addition to the program stars
observed with the HET, we have included in this study the spectra
of the Sun (Kurucz et al. 1984) and $\alpha$ CMi (Allende Prieto et al. 2002).

\section{Analysis}
\label{doit}

\subsection{Stellar parameters}

The effective temperature ($T_{\rm eff}$) has been estimated 
for all the program stars  
 by one or more authors using the Infrared Flux Method
(Blackwell \& Lynas-Gray 1994; Alonso et al. 1996, 1999). 
The $T_{\rm eff}$ of $\alpha$ Boo (Arcturus) 
has  been independently derived by 
di Benedetto (1998) and Griffin \& Lynas-Gray (1999). 
We have adopted the
values in Table 1. All our stars  were
analyzed by Allende Prieto \& Lambert (1999), and we accepted the 
surface gravities derived there from the (\bv) colors and the 
{\it Hipparcos} trigonometric parallaxes. 
The iron abundances in Table 1
are a straight average of previous spectroscopic analyses
compiled by Cayrel de Strobel et al. (2001).
A recent discussion of the preferred stellar parameters of $\alpha$ CMi 
can be found in Allende Prieto et al. (2002).
Fe\,{\sc i} lines at 4602.00 and 4602.94 \AA\ with accurate
$gf$-values and damping parameters were analyzed with
synthetic spectra computed from
flux-constant LTE model  atmospheres to estimate the projected
rotational velocities in Table 1.

\subsection{Measuring the line centers}

A variety of procedures  have been used previously to
determine the central wavelength of a spectral line. The most popular methods
involve some type of least-squares polynomial fitting or polynomial
interpolation (see, e.g., Neckel \& Labs 1990; Allende Prieto \& Garc\'{\i}a
L\'opez 1998).  
As the stellar spectral lines are asymmetric, 
 the details of the measuring procedure 
 affect the derived central wavelength.
Polynomial least-squares fits are well-suited for error estimation, 
however, there is some arbitrariness in the 
selection of the ideal order
and wavelength interval to use in the fits. For the slowly rotating stars 
analyzed here, we have found that a reasonable choice, the same for the 
whole sample, provides stable results regardless of the exact 
selection for the
order and wavelength interval.  Our 
particular choice  is to fit 35 m\AA\ around the line center with
a third order polynomial. Preceding the fitting by a cubic-spline interpolation
with a step of 5 m\AA\ proved useful for the HRS spectra, but degraded
slightly the quality of the results for the superior solar atlas.

To estimate the error in a 
derived central wavelength, we determine the horizontal
scatter between the polynomial fit F($\lambda$) and the 
observed spectrum $f(\lambda)$. For each observed wavelength $\lambda_i$ 
considered in the fit, we  find numerically the roots of the polynomial 
 $F(\lambda) - f(\lambda_i)$ using Laguerre's 
 method, select the appropriate root, 
 and then derive the rms scatter between the observed and fit wavelengths. 
 Assuming the errors are Gaussian, 
 this scatter is divided by the square root of the 
 number of points entering the polynomial fit to
 derive the uncertainty $\sigma_{\rm s}$. As an example, application 
 to the solar flux spectrum provides a nearly normal distribution 
 of uncertainties  that peaks at 0.00025 \AA\ and shows  
  bumps on the long-value tail, signaling blends. 
  Nave et al. (1994) classify their wavelengths into four 
categories,
depending on their uncertainty ($\sigma_{\rm l}$), which ranges 
from 0.4 m\AA\ to more than 
10 m\AA. Most lines have errors below 5 m\AA\ and about 
half of the 
lines have errors below 1 m\AA\footnote{This comparison indicates that an
extension and improvement of the laboratory wavelengths for neutral iron
 and other species is desirable. A spectrograph used
 to obtain stellar spectra may be well suited for this purpose 
 (see Allende Prieto 2001)}.
 
\begin{figure*}
\begin{center}
\includegraphics[width=4cm,angle=90]{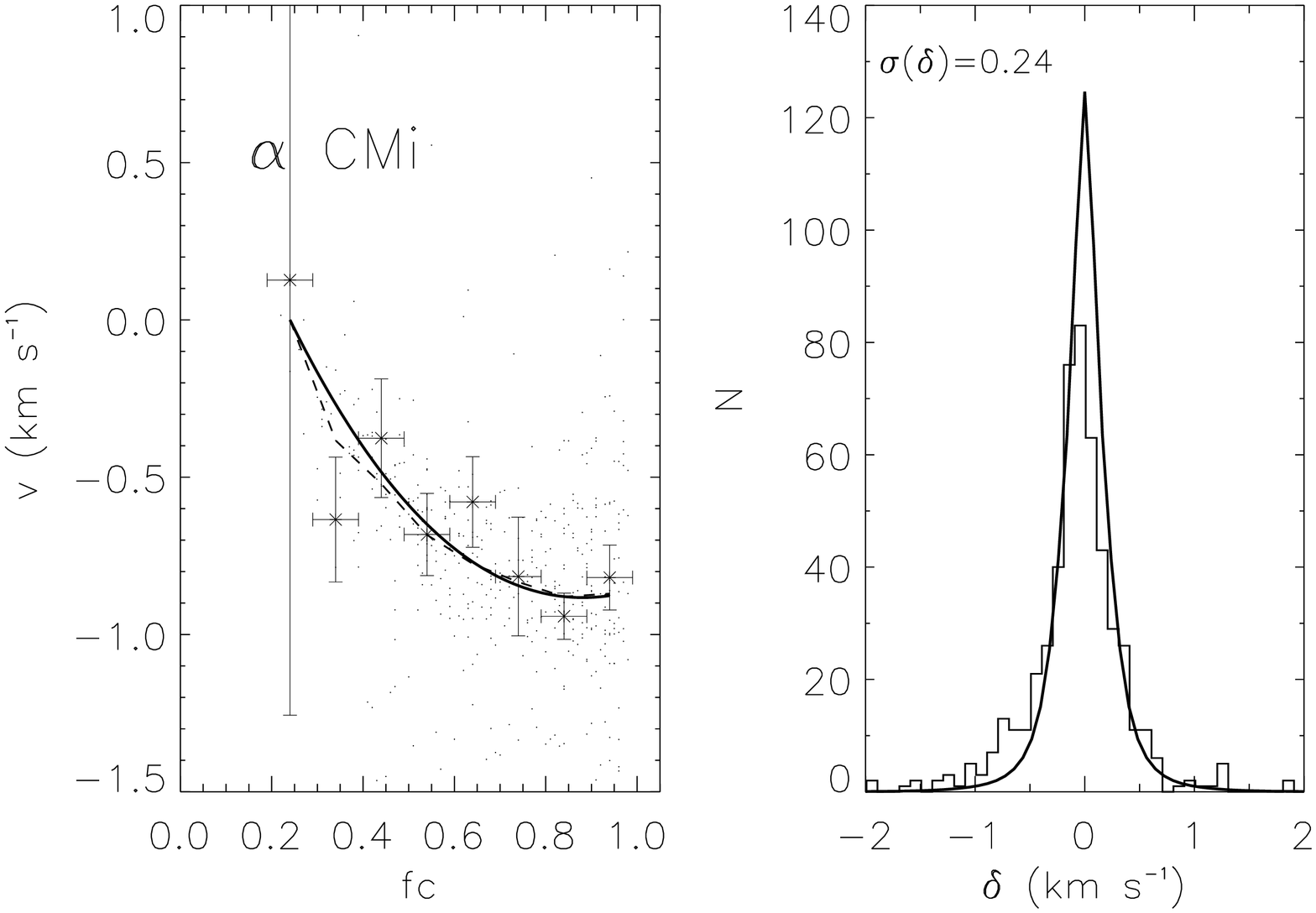}
\includegraphics[width=4cm,angle=90]{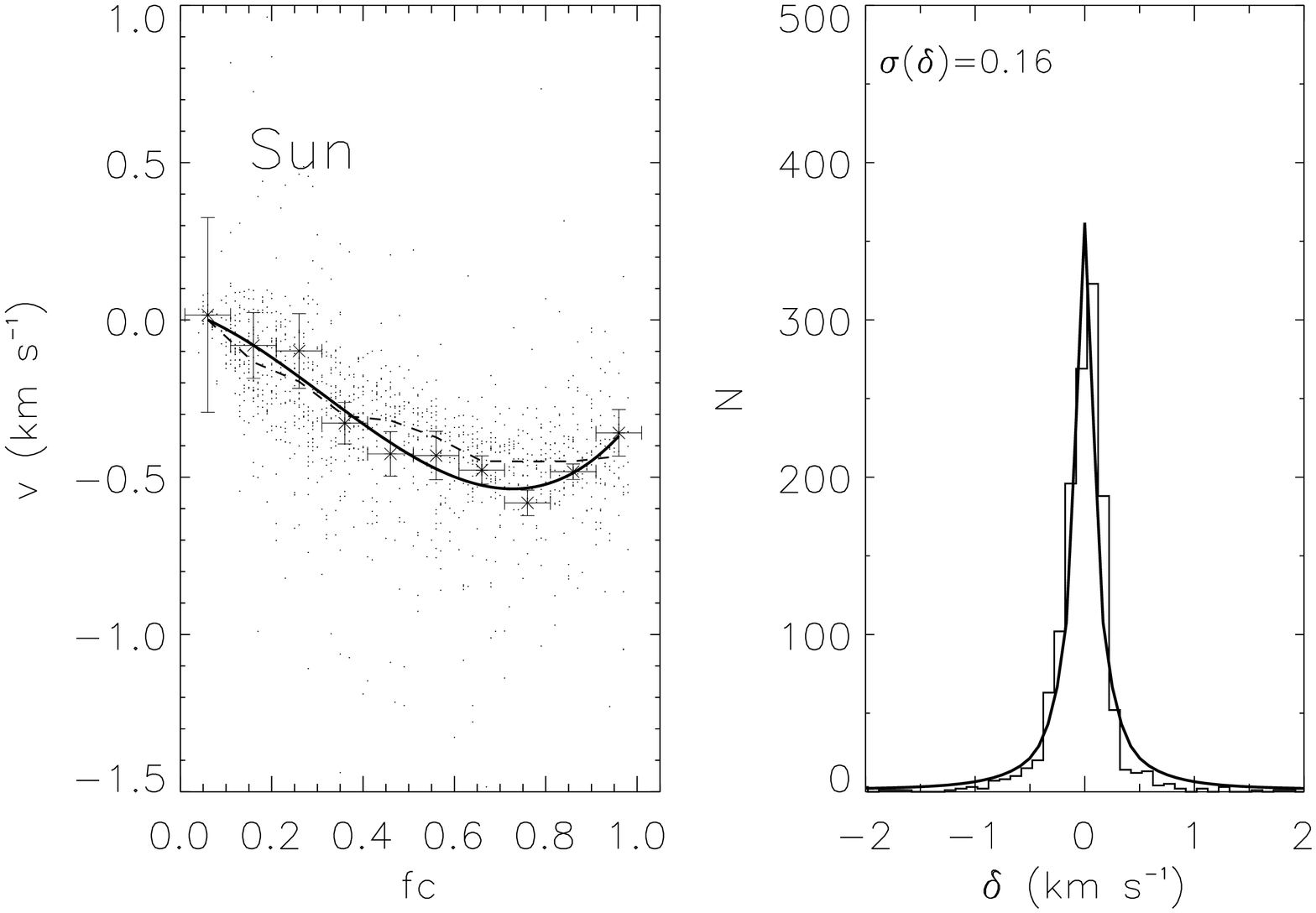}
\includegraphics[width=4cm,angle=90]{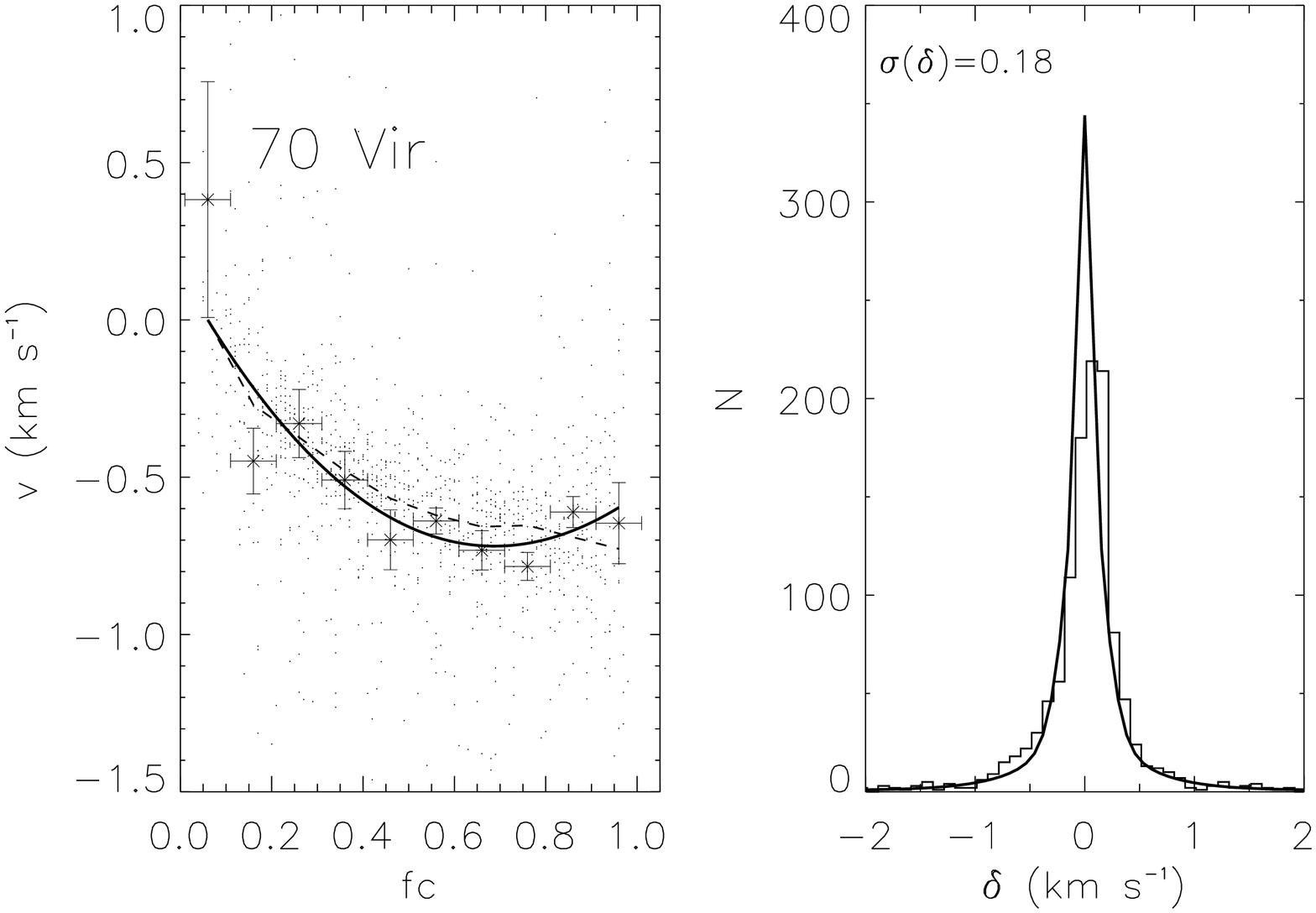}
\includegraphics[width=4cm,angle=90]{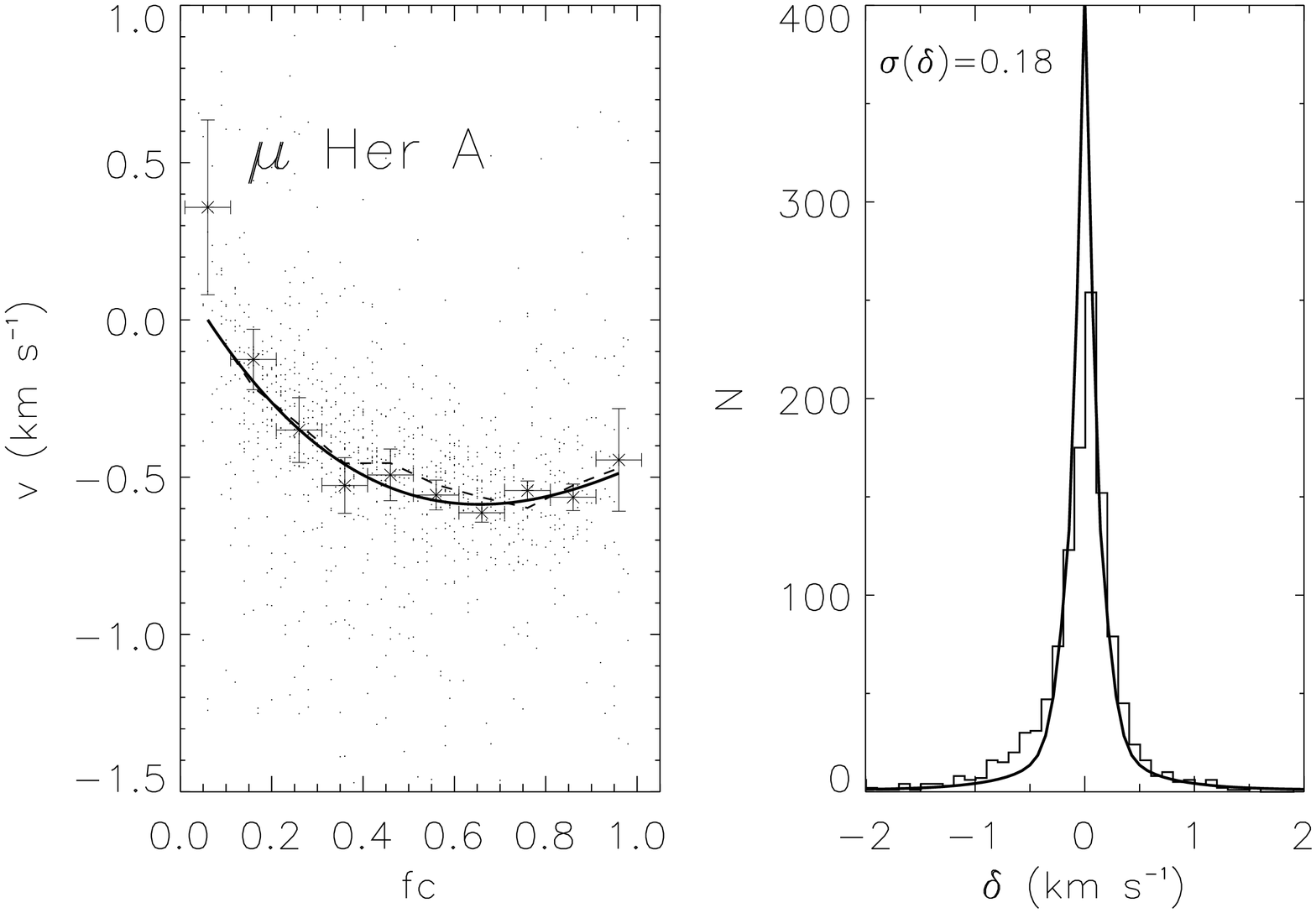}
\includegraphics[width=4cm,angle=90]{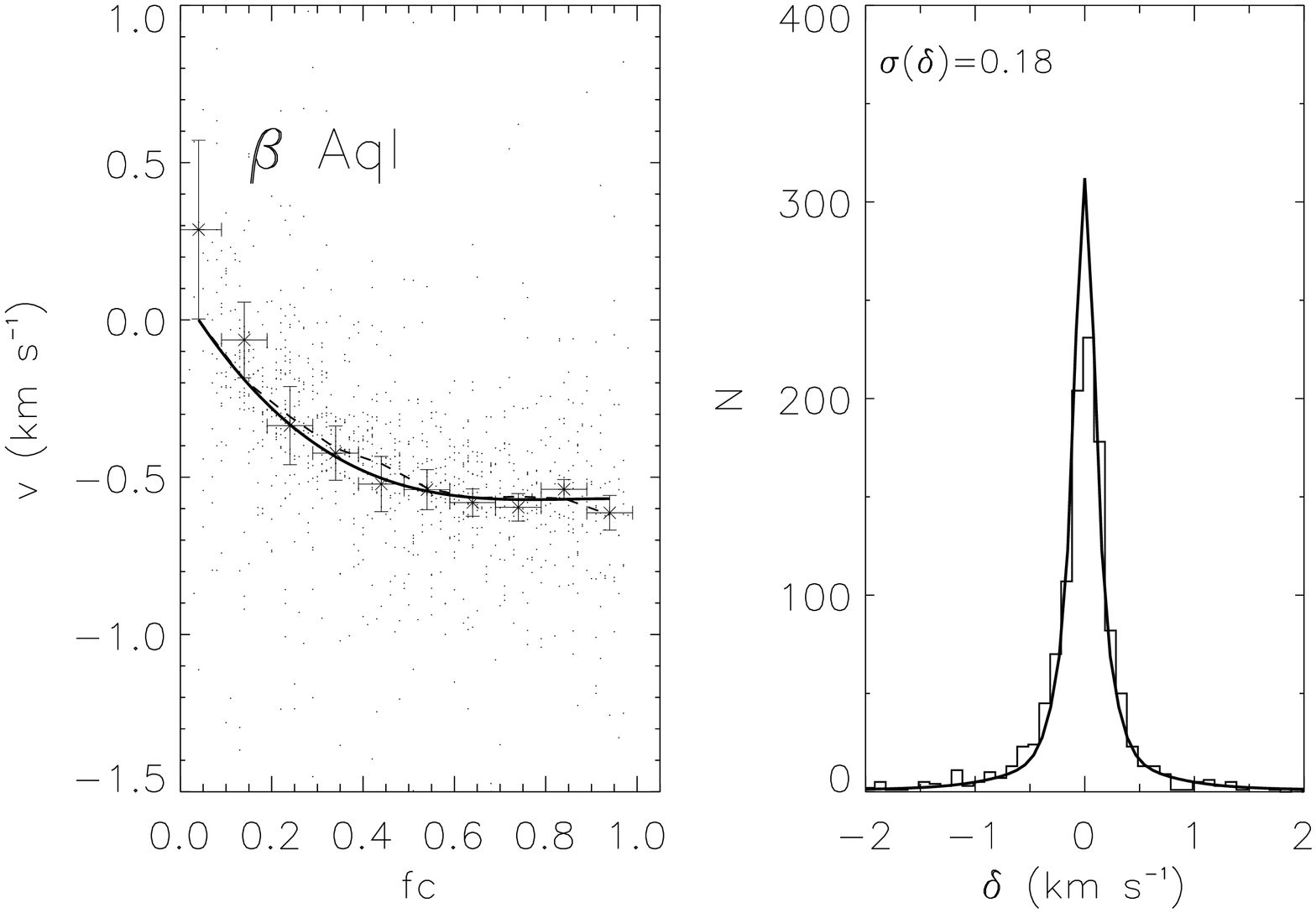}
\includegraphics[width=4cm,angle=90]{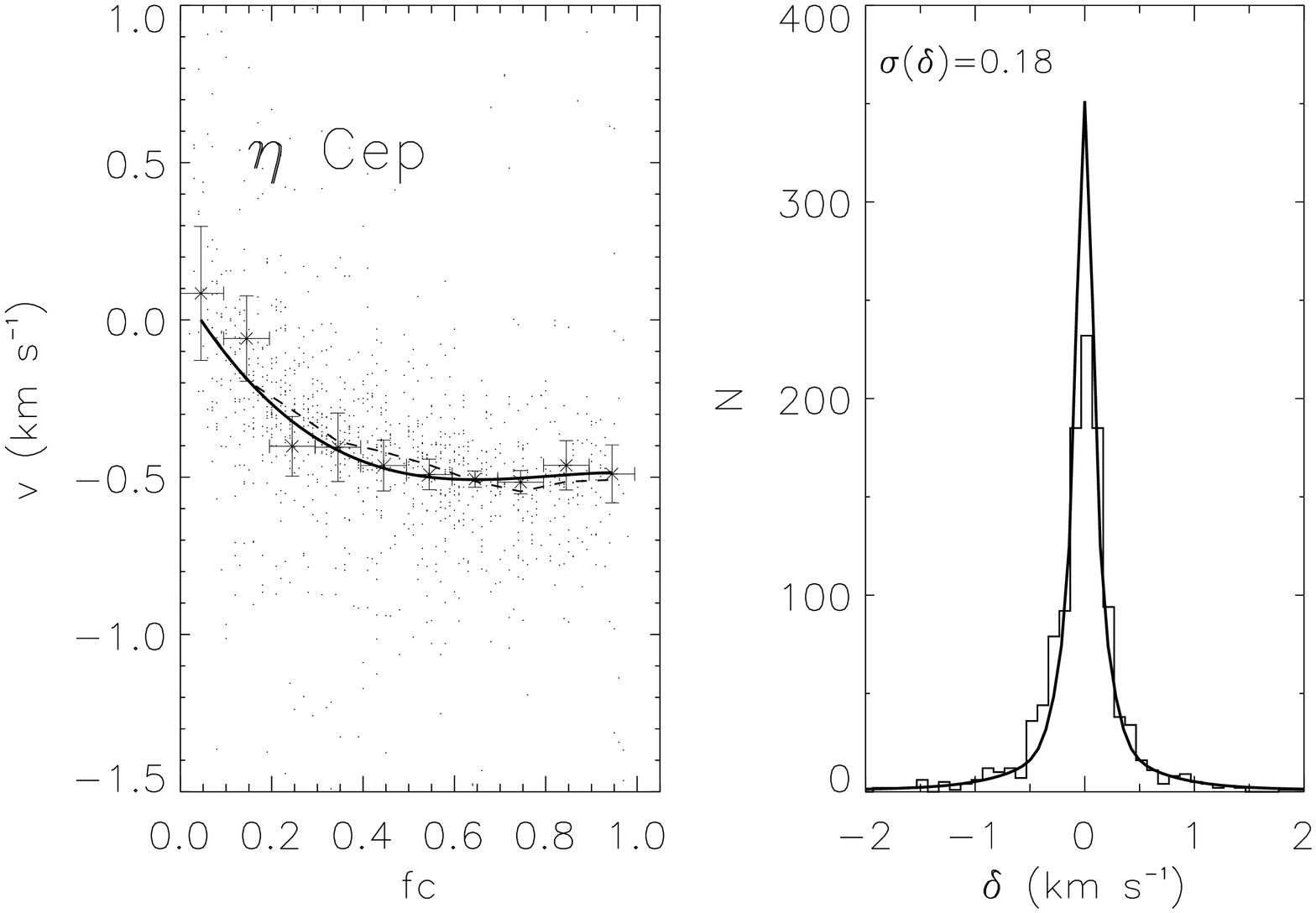}
\includegraphics[width=4cm,angle=90]{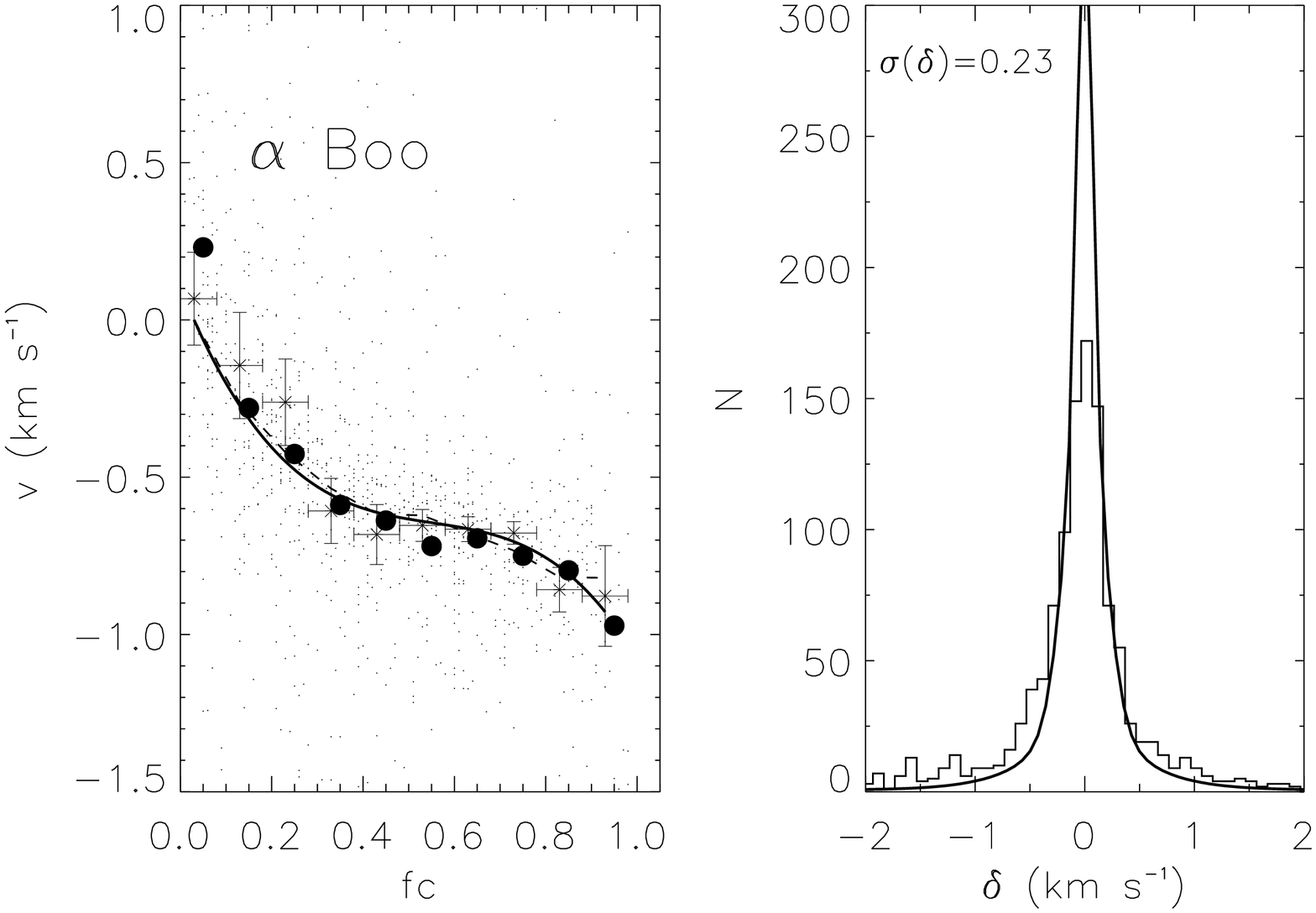}
\includegraphics[width=4cm,angle=90]{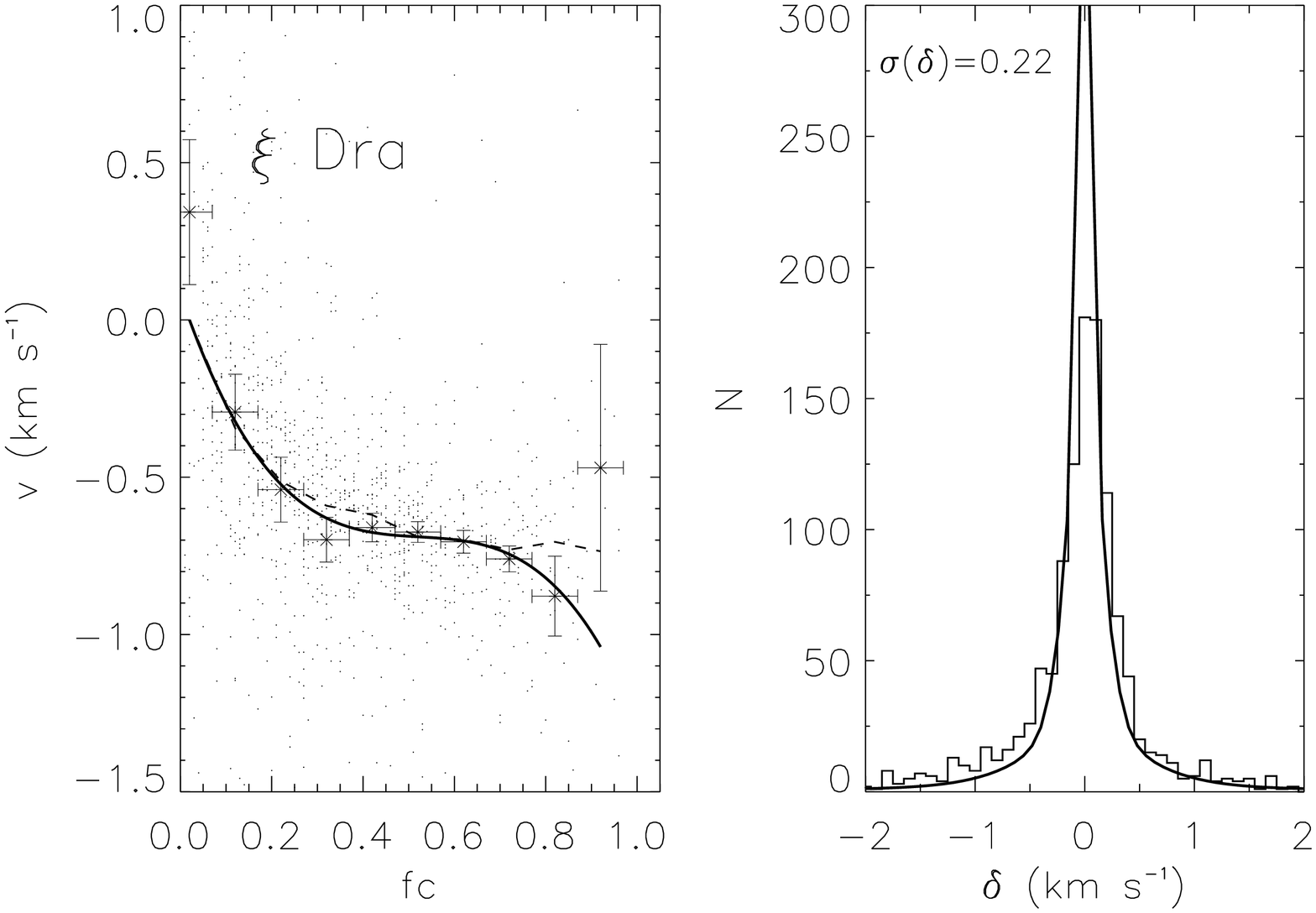}
\figcaption{
The left-hand panels show the relative velocity shifts of Fe {\sc i}
 lines with respect
to their rest wavelengths. The velocity scale for each star 
is shifted to have a null velocity
for the median shift of the strongest lines considered. The dots show the
shifts for the individual lines. The asterisks
 with error bars are mean values
after applying a median filter, and
the solid line is a polynomial fit to them. The dashed curve shows the median
shifts for the different bins. The right-hand panels
show histograms of the scatter of the velocity shifts around the polynomial
fit. The expected scatter (see text) is shown by the symmetric solid  curve. 
The mean velocity shifts determined from the analysis of the $\alpha$ Boo atlas
of Hinkle et al. (2000) are displayed as filled circles in the left-hand panel
for this star.
\label{shifts}}
\end{center}
\end{figure*} 

\tablenum{1}
\tablecolumns{7}
\tablewidth{0pc}
\begin{deluxetable}{cccccccc}
\tiny
\tablecaption{Observed sample \label{table1}}
\tablehead{
\colhead{Star} & \colhead{Obs. date} &  \colhead{Sp. Type} 
& \colhead{$T_{\rm eff}$\tablenotemark{a}} & \colhead{$\log g$} 
& \colhead{[Fe/H]}  &  \colhead{$v \sin i$}  & \colhead{$S$(v)} \\
 \colhead{}  & \colhead{} &   \colhead{}   &  \colhead{(K)} 
 &   \colhead{(dex)}  & \colhead{(dex)} & \colhead{(km s$^{-1}$)}  & \colhead{(km s$^{-1}$)}\\ }
\startdata
$\alpha$ CMi    & McDonald atlas & F5 IV & 6530 &    3.96 (0.02) &  $-$0.05 (0.03) & 2.8 (0.1)  	& 0.88 \\
Sun		& FTS atlas & G2 V  &	5770	&	4.44 (0.01) &   0.00	   & 1.9 (0.1) 		& 0.54 \\
70 Vir		& June 5  & G4 V    &	5455	&	4.04 (0.11)  &  $-$0.11\tablenotemark{b}        & 3.2 (0.1) 		& 0.72 \\
$\mu$ Her A	& May 28  & G5 IV   &	5523	&	3.87 (0.04)  &	0.23 (0.07) & 4.1 (0.2) 	& 0.59 \\
$\beta$ Aql	& June 11 & G8 IV   &	5040	&	3.56 (0.07)  &  $-$0.02 (0.10) & 3.1 (0.3) 	& 0.57 \\
$\eta$ Cep	& May 29  & K0 IV   &	4944	&	3.40 (0.18)  &  $-$0.12 (0.14) & 3.1 (0.1)	& 0.51 \\
\tableline
$\alpha$ Boo 	& June 9  & K1.5 III &	4290	&	2.00 (0.56)  &  $-$0.51 (0.03) & 3.0 (0.5)	& 0.93 \\
$\xi$ Dra	& May 29  & K2 III   &  4430    & 	2.31 (0.32)  &  $-$0.13 (0.04) & 5.1 (0.5)	& 1.04 \\
\enddata
\tablenotetext{a}{Uncertainties in $T_{\rm eff}$ are 100 K or smaller}
\tablenotetext{b}{A single measurement is available}
\end{deluxetable}

The uncertainties in the measured  wavelength shifts, 
$\sigma = \sqrt{\sigma_{\rm s}^2 + \sigma_{\rm l}^2}$, are then converted to
velocity, and their  distribution $n(\sigma)$ is determined. 
Then, we  use $n(\sigma)$ 
to predict the distribution of 
errors ($\delta$) in the observed velocity shifts  as

\begin{equation}
N(\delta) = \frac{1}{C} \int_{0}^{\infty}  \frac{n(\sigma)}{\sigma} 
\exp \left(-\frac{\delta^2}{2\sigma^2}\right) d\sigma,
\label{ndelta}
\end{equation}

\noindent where $C$ is chosen such  that 
$\int_{-\infty}^{\infty} N(\delta) d\delta = 1$. The derived $N(\delta)$ 
can be compared to the observations 
in order to constrain the intrinsic scatter. 
In all cases, the derived and measured histograms are similar.

\subsection{Searching for correlations}

The number of iron lines identified and measured for each star varies from
1340 to 1569\footnote{With the exception of $\alpha$ CMi, for which 
only 588 lines were measured}.
Solar analyses  have shown clear 
correlations of the line shifts with line strength  (Allende Prieto \& 
Garc\'{\i}a L\'opez 1998)
and wavelength (Hamilton \& Lester 1999). We explore here the variation
of the shifts with line strength, as this is the dominant effect. The line
equivalent width and the central line depth have been the most
widely used  line strength indicators. As pointed out by 
Pierce \& Lopresto (2000), 
the use of the equivalent width avoids saturation  
 for the strongest lines.
On the other hand, equivalent width  measurements are distorted 
by blends in the
line wings which may hardly affect line depths.
Blends are a  particularly serious problem
when dealing with cool stars, as some of those in our sample. Therefore, we 
have adopted the line depth.

The left-hand panels in Fig. \ref{shifts} show the line 
velocity shifts  for 
the stars in our sample as a function of the residual flux
at the center of the line $f_{\rm c}$ ($= 1 -$ line depth). In these panels, 
the shifts of the 
individual lines are marked with dots. Grouping the lines in bins of
0.1 in depth, and 
after applying a median filter, we derive the average values shown as
asterisks with error bars. The solid
 lines are 3$^{\rm th}$-order polynomial fits to
guide the eye, and the 
broken lines trace the original median values. 
We have not attempted to disentangle
the stellar radial velocities from the gravitational shifts and the 
 atmospheric motions.
The absolute values on the vertical
axis have been shifted to have a null median 
velocity shift for the strongest lines; a condition that 
the polynomial fits were as well forced to satisfy.
The right-hand panels show the scatter of the line shifts
about the average points. The scatter expected from Eq. (1), mainly the result
of the uncertainties in the laboratory wavelengths, the finite width of the 
lines, and the presence of photometric noise, is also displayed
 (thick symmetric curve). The $\sigma(\delta)$  
 given in the right-hand panels 
corresponds to a Gaussian  fit by least-squares 
to the observed scatter. The velocity span of the polynomial fits, 
$S$(v), is  listed in Table 1.

 A  spectroscopic 
atlas of $\alpha$ Boo at a superior resolving power and S/N  was 
published by Hinkle et al. (2000). The corresponding left-hand panel in 
 Figure \ref{shifts}  compares
the  average shifts from our measurements from the
atlas (filled circles) with those from 
the HRS spectrum (asterisks with error bars). The agreement is quite
satisfactory.  The $\sigma$ of the scatter in our spectrum and in the atlas
agree within 10 m s$^{-1}$.

\section{Discussion}
\label{end}

The pattern of the  velocity shifts is similar for all the stars,
 and in qualitative 
agreement with solar results. The net 
convective blue shifts of the lines strengthen toward deeper photospheric
layers, and therefore affect the weak lines more than the strong ones. 
Despite
our limited sample, there is an indication of a smooth dependence of
the average velocity shifts with spectral type. 
This effect  is clearer in our analysis of line shifts than 
in previous studies using bisectors. As argued in the introduction, this is
likely the result of line shifts measurements 
being less affected by blending than line bisectors in late-G and K type stars.

Our preliminary results indicate that the trend of  
line shifts as a function of line strength can
be determined as a function of spectral type and gravity. 
At least for late-type dwarfs, assuming the strongest lines in the 
spectrum are free from convective
shifts (as is the case for the solar photosphere), 
it is possible to  
correct for convective wavelength shifts to within $\simeq 0.2$ km s$^{-1}$.

The measured range of wavelength shifts for neutral iron lines is larger in 
$\alpha$ CMi than in cooler stars of classes IV and V.
 The range of wavelength shifts
is also larger,  reaching up to $\sim$ 1 km s$^{-1}$,  for the K giants 
than for the G-K dwarfs and subgiants in our
sample, which is not 
surprising. Although their lower
effective temperatures provide less flux to transport, their lower
atmospheric pressure implies larger convective velocities. 
 Fig. 1 suggests that, once we account for the observational errors,
  there is little room for intrinsic line-to-line scatter 
in the  F-G-K stars sampled here.
Convection is supposed to cease in main sequence
stars with spectral types earlier than about F2, but other velocity fields
 must
be present in those atmospheres, judging from the asymmetry of the spectral 
lines (e.g. Gray \& Nagel 1989). Determining accurate
 absolute radial velocities 
demands an understanding of the wavelength shifts in the 
spectra of these stars too.

Solar observations  have shown differences
in the center-to-limb variation of the granulation along the central 
meridian and the equator 
(e.g. Beckers \& Taylor 1980; 
Rodr\'{\i}guez Hidalgo, Collados, \& V\'azquez 1992). 
Observations of a large number of stars may
then show peculiarities at a given spectral type 
depending on the orientation of the rotational axis.  Stars with
enhanced  magnetic fields are also expected to be
peculiar in terms of observed wavelength shifts, as the enhanced magnetic
fields may hinder the convective motions. Furthermore, line shifts
may vary with time for stars exhibiting an analog of the solar
11 year cycle. 

Observations of line shifts for a significant number of
stars with high quality are required for a deeper understanding of 
granulation in stellar atmospheres, its relationship with other stellar
phenomena,  and the role of surface convection 
in the structure and evolution of stars. Our first results show that
measurements
of convective 
line shifts are a must in order to derive accurate absolute radial 
velocities.

We thank the HET staff for their outstanding job making possible science
observations with HRS since day one. 
The Hobby-Eberly Telescope is operated by McDonald Observatory on behalf
of The University of Texas at Austin, the Pennsylvania State University,
Stanford University, Ludwig-Maximilians-Universit\"at M\"unchen, and
Georg-August-Universit\"at G\"ottingen.
NSO/Kitt Peak FTS data used here were produced
by NSF/NOAO.
This research was supported in part by the NSF (grant AST-0086321).


\begin{thebibliography}{}

\bibitem[]{} Allende Prieto, C. 2001, 
The Spectrum of the Th-Ar Hollow-Cathode Lamp
Used with the 2dcoud\'e Spectrograph, 
{\tt http://hebe.as.utexas.edu/2dcoude/thar/} (astro-ph/0111172)

\bibitem[]{} Allende Prieto, C., Asplund, M., Garc\'{\i}a L\'opez, R. J., \& Lambert, D. L. 2002, \apj, in press (astro-ph/011055)

\bibitem[Allende Prieto \& Garcia Lopez(1998)]{1998A&AS..129...41A} Allende Prieto, C., \& Garc\'{\i}a L\'opez, R.~J.\ 1998, \aaps, 129, 41

\bibitem[Allende Prieto \& Lambert(1999)]{1999A&A...352..555A} Allende Prieto, C., \& Lambert, D.~L.\ 1999, \aap, 352, 555

\bibitem[Alonso, Arribas, \& Martinez-Roger(1996)]{1996A&A...313..873A} Alonso, A., Arribas, S., \& Mart\'{\i}nez-Roger, C.\ 1996, \aap, 313, 873

\bibitem[Alonso, Arribas, \& Mart{\' i}nez-Roger(1999)]{1999A&AS..140..261A} Alonso, A., Arribas, S., \& Mart\'{\i}nez-Roger, C.\ 1999, \aaps, 140, 261

\bibitem[Asplund et al.(2000)]{2000A&A...359..729A} Asplund, M., Nordlund, {\AA}., Trampedach, R., Allende Prieto, C., \& Stein, R.~F.\ 2000, \aap, 359, 729

\bibitem[Beckers \& Taylor(1980)]{1980SoPh...68...41B} Beckers, J.~M.,~\& Taylor, W.~R.\ 1980, \solphys, 68, 41

\bibitem[Blackwell \& Lynas-Gray(1994)]{1994A&A...282..899B} Blackwell, D.~E.,~\& Lynas-Gray, A.~E.\ 1994, \aap, 282, 899

\bibitem[Cayrel de Strobel, Soubiran, \& Ralite(2001)]{2001A&A...373..159C} Cayrel de Strobel, G., Soubiran, C., \& Ralite, N.\ 2001, \aap, 373, 159

\bibitem[di Benedetto(1998)]{1998A&A...339..858D} di Benedetto, G.~P.\ 1998, \aap, 339, 858

\bibitem[Dravins(1974)]{1974A&A....36..143D} Dravins, D.\ 1974, \aap, 36, 143

\bibitem[Dravins(1985)]{1985srv..proc..311D} Dravins, D.\ 1985, IAU Colloq.~88: Stellar Radial Velocities, 311

\bibitem[Dravins(1987)]{1987A&A...172..211D} Dravins, D.\ 1987, \aap, 172, 211

\bibitem[Dravins(1999)]{1999psrv.conf..268D} Dravins, D.\ 1999, ASP Conf.~Ser.~185: IAU Colloq.~170: Precise Stellar Radial Velocities, 268

\bibitem[Dravins \& Nordlund(1990)]{1990A&A...228..203D} Dravins, D.,~\& Nordlund, \AA.\ 1990, \aap, 228, 203

\bibitem[Gray(1980)]{1980ApJ...235..508G} Gray, D.~F.\ 1980, \apj, 235, 508

\bibitem[Gray(1981)]{1981ApJ...251..583G} Gray, D.~F.\ 1981, \apj, 251, 583

\bibitem[Gray \& Nagel(1989)]{1989ApJ...341..421G} Gray, D.~F.,~\& Nagel, T.\ 1989, \apj, 341, 421

\bibitem[Gray \& Toner(1986)]{1986PASP...98..499G} Gray, D.~F.,~\& Toner, C.~G.\ 1986, \pasp, 98, 499

\bibitem[Griffin \& Lynas-Gray(1999)]{1999AJ....117.2998G} Griffin, R.~E.~M.,~\& Lynas-Gray, A.~E.\ 1999, \aj, 117, 2998

\bibitem[Hamilton \& Lester(1999)]{1999PASP..111.1132H} Hamilton, D.,~\& Lester, J.~B.\ 1999, \pasp, 111, 1132

\bibitem[Hinkle, Wallace, Valenti, \& Harmer(2000)]{2000vnia.book.....H} Hinkle, K., Wallace, L., Valenti, J., \& Harmer, D.\ 2000, 
Visible and Near Infrared Atlas of the Arcturus Spectrum 3727-9300 
\AA\  (San Francisco: ASP)

\bibitem[Kurucz, Furenlid, \& Brault(1984)]{1984sfat.book.....K} Kurucz, R.~L., Furenlid, I., \& Brault, J.\ 1984, National Solar Observatory Atlas (Sunspot, New Mexico: NSO)

\bibitem[Lindegren, Dravins, \& Madsen(1999)]{1999psrv.conf...73L} Lindegren, L., Dravins, D., \& Madsen, S.\ 1999, ASP Conf.~Ser.~185: IAU Colloq.~170: Precise Stellar Radial Velocities, 73


\bibitem[Nave et al.(1994)]{1994ApJS...94..221N} Nave, G., Johansson, S., Learner, R.~C.~M., Thorne, A.~P., \& Brault, J.~W.\ 1994, \apjs, 94, 221

\bibitem[Neckel \& Labs(1990)]{1990SoPh..126..207N} Neckel, H.,~\& Labs, D.\ 1990, \solphys, 126, 207

\bibitem[Pierce \& Lopresto(2000)]{2000SoPh..196...41P} Pierce, A.~K.,~\& Lopresto, J.~C.\ 2000, \solphys, 196, 41

\bibitem[Rodriguez Hidalgo, Collados, \& Vazquez(1992)]{1992A&A...264..661R} Rodr\'{\i}guez Hidalgo, I., Collados, M., 
\& V\'azquez, M.\ 1992, \aap, 264, 661

\bibitem[Tull(1998)]{1998SPIE.3355..387T} Tull, R.~G.\ 1998, \procspie, 3355, 387

\end{thebibliography}
\end{document}